\documentstyle[twocolumn,graphicx,aps]{revtex}

\begin{document}

\draft

\title{Ferromagnetism in a lattice of Bose condensates}

\author{Han Pu, Weiping Zhang, and Pierre
Meystre}
\address{
{Optical Sciences Center, The University of Arizona,
Tucson, AZ 85721} \\ (\today)
\\ \medskip}\author{\small\parbox{14.2cm}{\small\hspace*{3mm}
We show that an ensemble of spinor Bose-Einstein condensates
confined in a one dimensional optical lattice can undergo a
ferromagnetic phase transition and spontaneous
magnetization arises due to the magnetic dipole-dipole interaction. This
phenomenon is analogous to ferromagnetism in solid state physics,
but occurs with bosons instead of fermions.\\
\\[3pt]PACS numbers: 03.75.Fi, 75.45.+j, 75.60.Ej}} \maketitle

\narrowtext

The Heisenberg model of spin-spin interactions is considered as
the starting point for understanding many complex magnetic
structures in solids. In particular, it explains the existence of
ferromagnetism and antiferromagnetism at temperatures below the
Curie temperature. It is defined by the spin Hamiltonian
\cite{huang}
\[
H_{{\rm spin}}=-\sum J_{ij}\, {\bf S}_i \cdot {\bf S}_j,
\]
where ${\bf S}_i$ is the spin operator for $i$-th electron, and
$J_{ij}$ are known as the exchange coupling constants. This
Hamiltonian arises from the direct Coulomb interaction among
electrons and the Pauli exclusion principle. In addition to the
exchange interaction, there exists another important type of
magnetic interaction, the magnetic dipole-dipole interaction.
However, in solid materials the dipolar coupling is typically
several orders of magnitude weaker than the exchange coupling, and
would correspond to Curie temperatures much below the observed
ones. Hence its contribution to the spin Hamiltonian can be
neglected in practice.\footnote{We note however that the magnetic
dipole-dipole interaction plays an important role in domain
formation in macroscopic samples.} It follows from this argument
that ferromagnetism is not generally expected to occur in bosonic
lattices of neutral atoms, a result of the inapplicability of the
Pauli principle, the absence of Coulomb interaction and small
atomic magnetic dipole moments. However, we qualify this remark by
noting that as a result of accurate recent measurements of the
scattering length of spin-changing collisions, it is now
established that the ground state of optically trapped $^{87}$Rb
spinor Bose condensates is ferromagnetic, an important result that
we use in the following \cite{heinzen,green}.

The goal of this paper is to show that this result, combined with
the recent experimental realization of regular arrays of
Bose-Einstein condensates in optical lattices, leads to a
situation where it becomes possible to carry out detailed static
and dynamic studies of magnetism on one to three-dimensional
periodic lattices.

Quantum degenerate Bose gases on optical lattices were first used
to demonstrate ``mode-locked'' atom lasers \cite{yale1}. It has
also been theoretically demonstrated that they undergo a Mott
insulator phase transition as the depth of the lattice wells is
increased \cite{zoller}. Recently, they have become of interest in
the study of quantum chaos \cite{raizen,carretero}. Here, we show
that spinor condensates, localized in optical lattices deep enough
for the individual sites to be independent, can undergo a
ferromagnetic-like phase transition that leads to a
``macroscopic'' magnetization of the condensate array.

We consider specifically the case of spinor $^{87}$Rb condensates
\cite{mit}, which are as we have mentioned individual ferromagnets
of random directions in the absence of external fields and
magnetic dipole-dipole interaction. We show that the magnetic
dipole-dipole interaction between lattice sites can spontaneously
align the magnetization of the individual sites. This is possible
because of the Bose enhanced magnetic dipole moments of the
condensate which in turn enhances the strength of the magnetic
dipolar interaction.

Our starting point is the Hamiltonian $H$ describing an $F=1$
spinor condensate at zero temperature trapped in an optical
lattice, subject to a magnetic dipole-dipole interaction $H_{dd}$
and is coupled to an external magnetic field via the magnetic
dipole Hamiltonian $H_B$\cite{zhang,ho2,ohmi,law},
\begin{equation}
H = H_0 + H_{dd} + H_B.
\end{equation}
Here
\begin{eqnarray*}
H_0& =& \int d^3r\, \psi_\alpha^{\dagger}({\bf r}) [-\hbar^2
\nabla^2 /2m + V_L({\bf r}) ]\psi_\alpha({\bf r})\\ &+&(
\lambda_s/2)\int d^3r\, \psi_\alpha^{\dagger}({\bf r})
\psi_\beta^{\dagger}({\bf r}) \psi_\beta ({\bf r})
\psi_\alpha({\bf r}) \\ &+& (\lambda_a/2)\int
d^3r\,\psi_\alpha^\dagger({\bf r}) \psi_\mu^\dagger({\bf r}) {\bf
F}_{\alpha \beta} \cdot {\bf F}_{\mu \nu} \psi_\nu({\bf
r})\psi_\beta ({\bf r})
\end{eqnarray*}
describes the interaction of the atoms with the lattice potential
$V_L({\bf r})$ and ground state collisions. It includes an
implicit sum over the indices \{$\alpha,\beta,\mu,\nu$$\} = \{$0,
$\pm 1$\} that label the three Zeeman sublevels. The parameters
$\lambda_s$ and $\lambda_a$ characterize the short-range
spin-independent and spin-changing $s$-wave collisions,
respectively. Specifically, $\lambda_a$ is proportional to the
difference between the $s$-wave scattering lengths in the triplet
and singlet channels\cite{ho2,ohmi,law}. It has recently been
measured to be negative for $^{87}$Rb, accounting for its
ferromagnetic ground state \cite{heinzen}.

Our model includes the long-range magnetic dipole-dipole
interaction between different lattice sites, but neglect it within
each site, assuming that it is much weaker than the $s$-wave
interaction described by $H_0$. We also assume that the optical
lattice potential is deep enough that there is no spatial overlap
between the condensates at different lattice sites. We can then
expand the atomic field operator as $\psi({\bf r})=\sum_i
\sum_{\alpha=0, \pm 1} \hat{a}_\alpha(i) \phi_i({\bf r})$ where
$i$ labels the lattice sites.

The Hartree wave function $\phi_i({\bf r})$, determined by
minimizing the total energy, is the wave function of the
condensate at the $i$-th site. It is assumed that all Zeeman
components share the same spatial wave function. If the
condensates at each lattice sites contain the same number $N$ of
atoms, then the ground-state wave functions for different sites
have the same form $\phi_i({\bf r}) =\phi({\bf r}-{\bf r_i})$.
Under this condition, the dipolar interaction potential is
$H_{dd} = \sum_{i,j \neq i}V_{dd}^{ij}$ with
\begin{equation}
V_{dd}^{ij}  =
\frac{\mu_0}{4\pi} \left[\frac{\vec{\mu}_i \cdot  \vec{\mu}_j}{|{\bf
r}_{ij}|^3}-\frac{3(\vec{\mu}_i \cdot \hat{{\bf r}}_{ij}) ( \vec{\mu}_j \cdot
\hat{{\bf r}}_{ij})}{|{\bf r}_{ij}|^3}  \right] ,
\end{equation}
where $\mu_0$ is the vacuum permeability, ${\bf r}_{ij}={\bf
r}_i-{\bf r}_j$, $\hat{{\bf r}}_{ij} = {\bf r}_{ij}/ | {\bf
r}_{ij}|$, ${\bf r}_i$ is the coordinate of the $i$-th site, and
$\vec{\mu}_{i}=\gamma {\bf S}_{i}$ is the magnetic dipole moment
at site $i$, with ${\bf S}_{i}=\hat{a}^{\dagger}_{\alpha}(i) {\bf
F}_{\alpha \beta} \hat{a}_{\beta}(i)$ being the total angular
momentum operator and $\gamma=g_F \mu_B$ the gyromagnetic ratio.
Finally, the coupling of the atoms to the external magnetic field
${\bf B}_{\rm ext}$ is described by
$$H_B = -\gamma \sum_i {\bf
S}_i \cdot {\bf B}_{\rm ext}. $$

In this letter we consider a one-dimensional optical lattice along
the $z$-direction, which we also choose as the quantization axis.
Hence
\[ V_{dd}^{ij}
 = \frac{\mu_0}{4\pi |{\bf r}_{ij}|^3} \left[\vec{\mu}_i
\cdot  \vec{\mu}_j-3{\mu}_i^z
 {\mu}_j^z  \right], \]
with $\mu_j^z$ being the $z$-component of $\vec{\mu}_j$. In the
absence of long-range magnetic dipole-dipole interaction and of
external magnetic fields, the individual condensates can therefore
be considered as independent ``magnets'' whose pseudo-spin vectors
point in random directions, with no spin correlations between
sites. Our goal is to determine the spin structure of the system
if the different sites are allowed to interact with each other
through the magnetic dipole-dipole interaction.

In the absence of spatial overlap between individual condensates,
and neglecting unimportant constants, the total Hamiltonian of the
system takes the form \cite{law}
\begin{eqnarray}
H=&&\sum_i \left[ \lambda_a' {\bf S}_i^2  +  \gamma \sum_{j \neq i} \lambda_{ij}
{\bf S}_i \cdot {\bf S}_j \right. \nonumber \\
&& \left. - 3\gamma \sum_{j \neq i} \lambda_{ij} S_i^z
S_j^z  - \gamma {\bf S}_i \cdot {\bf
B}_{\rm ext}\right], \label{h1}
\end{eqnarray}
where $\lambda_a'=(1/2)\lambda_a \int d^3r \, |\phi_i ({\bf
r})|^4$ and
$
\lambda_{ij} = \gamma \mu_0/(4\pi|{\bf r}_{ij}|^3).
$

In general the external magnetic field consists of two
contributions: a controlled, external applied field; and an
effective ``stray'' field that accounts for all possible effects
from the experimental environment and the system itself. A typical
example is the environmental magnetic fluctuations. In the present
case, we take the applied field along the quantization axis $z$.
The effective environmental magnetic field, in contrast, can have
any orientation. We choose it without loss of generality to be
along the transverse direction, including any longitudinal
component in the definition of the applied field. Hence the
external field has the form
\[ {\bf B}_{\rm ext}=B_z \hat{{\bf z}} + B_{\rho} \hat{{\bf \rho}}, \]
where $\rho=\sqrt{x^2+y^2}$ is the radial coordinate.

If the optical lattice is sufficiently long, one can safely
neglect the boundary effects and concentrate on a generic single
site $i$ of spin ${\bf S}$. Its Hamiltonian reads
\begin{eqnarray}
h=&& \lambda_a' {\bf S}^2 - \gamma {\bf S} \cdot \left[ \left(B_z+2 \sum_{j \neq
i} \lambda_{ij} S_j^z \right) \hat{\bf z} \right. \nonumber \\
&& + \left. \left(B_{\rho} -
\sum_{j \neq i} \lambda_{ij} S_j^{\rho}  \right) \hat{\bf \rho}\right] .
\label{hi}
\end{eqnarray}

We determine its ground state in the mean-field approximation
(also known as the Weiss molecular potential approximation)
\cite{solid}. It consists in replacing the operators
$S_j^{\alpha}$, $\alpha =\rho, z$, by their ground-state
expectation value
\begin{equation}
\langle S_j^{\alpha} \rangle \rightarrow M_{\alpha} =N m_{\alpha}
,
\end{equation}
which is assumed to be the same for different sites. We remark
that $m_z$ is nothing but the difference in population of the
Zeeman sublevels of magnetic quantum numbers $\pm 1$. This allows
us to approximate the Hamiltonian (\ref{hi}) by
\begin{equation}
h_{\rm {mf}} =  \lambda_a' {\bf S}^2- \gamma {\bf S} \cdot {\bf
B}_{{\rm eff}}, \label{heff}
\end{equation}
where we have introduced the effective magnetic field
\begin{equation}
{\bf B}_{\rm {eff}} = (B_z+2\Lambda m_z) \hat{\bf z} + (B_x -
\Lambda m_{\rho}) \hat{\bf \rho}
\end{equation}
and $ \Lambda = N \sum_{j\neq i} \lambda_{ij} $.

We mentioned that the individual spinor condensates at the lattice
sites are ferromagnetic, $\lambda_a' <0$.  In that case, the
ground state of the mean-field Hamiltonian (\ref{heff}) must
correspond to a situation where the condensate at the site $i$
under consideration must be aligned along ${\bf B}_{{\rm eff}}$
and takes its maximum possible value $N$. That is, the ground
state of the mean-field Hamiltonian (\ref{heff}) is simply
\begin{equation}
|GS \rangle = |N, N \rangle_{{\bf B}_{{\rm eff}}}, \label{gs}
\end{equation}
where the first number denotes the total angular momentum and the
second its component along the direction of ${\bf B}_{{\rm eff}}$.
Note that $|GS \rangle$ represents a spin coherent state in the
basis of $|S,S_z \rangle$. The fact
that the ground state magnetic dipole moment of each lattice site
is $N$ times that of an individual atom results in a significant
magnetic dipole-dipole interaction even for lattice points
separated by hundreds of nanometers. This feature, which can be
interpreted as a signature of Bose enhancement, is in stark
contrast with usual ferromagnetism, where the magnetic interaction
is negligible compared to exchange and where the use of fermions
is essential\cite{solid}.

The mean-field ground state of Eq. (\ref{gs}) allows us to
calculate the magnetization components $m_z$ and $m_x$. One finds
readily
\begin{equation}
\label{mag}
m_\alpha = \frac{1}{N} \langle GS | S_i^\alpha |
GS  \rangle =  \cos \theta_\alpha ,
\end{equation}
where $\theta_\alpha$ is the
angle between ${\bf B}_{{\rm eff}}$ and the $\alpha$-axis.

For $B_z=0$, Eq.~(\ref{mag}) yields
\begin{mathletters}
\label{mxz}
\begin{eqnarray}
m_z &=& \frac{2 \Lambda m_z}{\sqrt{(2 \Lambda m_z)^2+
(B_{\rho}- \Lambda m_{\rho})^2}},
 \\
m_{\rho} &=& \frac{B_{\rho}-\Lambda m_{\rho}}{\sqrt{(2 \Lambda m_z)^2+(B_{\rho}-
\Lambda m_{\rho})^2}} .
\end{eqnarray}
\end{mathletters}
The solutions to Eqs.~(\ref{mxz}) can be divided into two cases:
\begin{enumerate}
\item
For $B_{\rho} \ge 3\Lambda$, the only solutions are $m_z=0$ and
$m_{\rho}=1$. That is, the lattice of condensates is magnetically
polarized along the transverse magnetic field.
\item For $B_{\rho} < 3\Lambda$, there are two coexisting sets of
solutions: (i) $m_z=0$ and $m_{\rho}=1$; and (ii) $m_z=\pm
\sqrt{1-(B_{\rho}/ 3\Lambda)^2}$ and $m_{\rho}=B_{\rho}/3\Lambda$.
However, it is easily seen that the state associated with the
latter solutions has the lower energy. Hence it corresponds to the
true ground state, while solution i) represents an unstable
equilibrium.
\end{enumerate}

We have, then, the following situation: As the environmental
effective magnetic field strength is reduced below a critical
value $3 \Lambda$, the lattice ceases to be polarized along the
direction of that field. A phase transition occurs, and a {\em
spontaneous magnetization} along the $z$-direction appears,
characterized by a finite $m_z$.

This phenomenon is reminiscent of conventional ferromagnetism.
Indeed, our model is somewhat analogous to the Ising
model\cite{huang}, with the environmental transverse magnetic
field $B_{\rho}$ playing the role of temperature. For $B_{\rho}=0$
--- corresponding to zero temperature in Ising model --- the spins
at each lattice site ${\bf S}_i$ align themselves along the
lattice direction, even in the absence of longitudinal field. This
spontaneous spin  magnetization diminishes as $B_{\rho}$
increases, and completely vanishes if $B_{\rho}$ exceeds the
critical value $3\Lambda$ --- the analog of the Curie temperature
in the Ising model. We note however that the two cases exhibit
important qualitative differences: For example, no spontaneous
magnetization occurs in 1D Ising model, for any finite
temperature\cite{huang}.

\begin{figure}
\begin{center}
    \includegraphics*[width=0.72\columnwidth,
height=0.5\columnwidth]{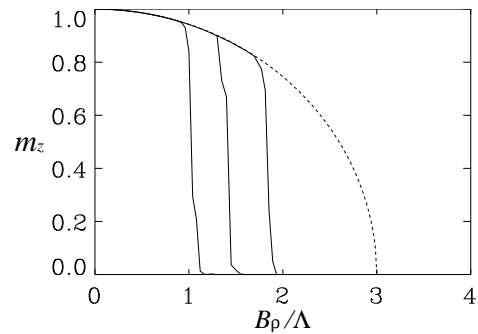} \vspace{3 mm} \caption{Spontaneous
magnetization as a function of environmental magnetic field strength.
Only the positive values are plotted. The dashed line represents
the mean-field result $m_z=\sqrt{1-(B_x/3\Lambda)^2}$ and the
solid lines, from left to right, correspond to the exact numerical
results for a two-site lattice with $N=10$, 15 and 25 atoms,
respectively.} \label{fig1} \end{center}
\end{figure}

Our analysis so far assumes an infinite one-dimensional lattice.
This assumption is required for the validity of mean-field
approximation. We note however that the appearance of a
spontaneous magnetization does not rely on this condition being
fulfilled. We demonstrate this point by considering just two
lattice sites, that is, a double-well situation. In this case, we
can numerically solve the Hamiltonian (\ref{h1}), without invoking
the mean-field approximation, by expanding the Hamiltonian matrix
on the basis of $|S_1=N, S_1^z \rangle \otimes |S_2=N, S_2^z
\rangle$. Figure~\ref{fig1} shows the spontaneous magnetization
$m_z$ as a function of the external field strength $B_{\rho}$. As
the Hamiltonian matrix has a size of $(2N+1)^2 \times (2N+1)^2$,
we cannot use a very large $N$. However, as we can see from the
figure, as $N$ increases, $m_z$ rapidly approaches the mean-field
results. Hence the mean-field treatment is in fact a surprisingly
good approximation for our purpose.

To estimate the feasibility of an experimental detection of the
spontaneous magnetization, we consider as an example the $F=1$
electronic ground state ($^3$S$_{1/2}$) of $^{87}$Rb, for which
$\lambda_a' < 0$\cite{heinzen,green}. The Land\'{e} factor for
this state is $g_F=-1/2$. For an optical potential of period equal
to 426 nm (nearest neighbor separation) we find $ \Lambda = N
\sum_{j \neq i} \lambda_{ij} \approx 1.6N \times 10^{-7} \;\;{\rm
G}$. If the particle number at each site is $N=2000$, this gives a
critical value of the environmental magnetic field of $3\Lambda =
1$mG. These numbers indicate that the observation of spontaneous
magnetization in a lattice of spinor condensates is well within
experimental reach.

\begin{figure}  \begin{center}
\includegraphics*[width=0.72\columnwidth,  height=0.5\columnwidth]{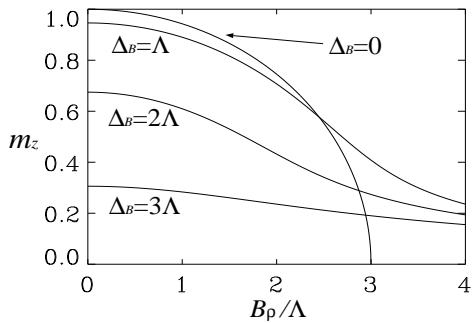}
\vspace{3 mm} \caption{Mean-field results of the time-averaged
spontaneous  magnetization as a function of environmental magnetic
field strength, taking the magnetic field fluctuations into
accout. Only the positive values are plotted.} \label{fig2}
\end{center}  \end{figure}

In a typical experimental situation, the environmental magnetic
field is likely to exhibit temporal fluctuations. To account for
them, we assume that this field has a uniform angular distribution
and that the fluctuations in field strength have a width
$\Delta_B$. The time-averaged environmental field is again assumed
to be along the transverse direction, with strength $B_{\rho}$.
Assuming the magnetization of the system follows the field
fluctuations adiabatically, it is not difficult to include this
effect into our mean-field treatment. The time-averaged
spontaneous magnetization along $z$-axis under satisfies then the
cubic equation
\[ (1-m_z^2)(9 \Lambda^2 m_z^2+ \Delta_B^2/2)=(B_{\rho}^2+\Delta_B^2) m_z^2. \]
Figure~\ref{fig2} illustrates $m_z$ as a function of $B_{\rho}$
for various $\Delta_B$. As we can see, the effect of the field
fluctuation is to decrease the maximum spontaneous magnetization,
and to increase the critical field strength at which the
spontaneous magnetization vanishes. The maximun $m_z$ at
$B_{\rho}=0$ decreases from 1 to 1/3 when $\Delta_B$ increases
from 0 to infinity.

Recently, condensates trapped in periodic optical potentials have
attracted much attention. The phase coherence and atom statistics
of the system have been studied
experimentally\cite{yale1,yale2,hansch}. However their magnetic
properties have not been fully explored. A few years ago, Meacher
{\em et al.} observed experimentally the paramagnetic behavior of
cold (but non-condensed) cesium atoms confined in an optical
lattice\cite{meacher}. Here we have demonstrated that by replacing
the cold atoms with a spinor condensate, ferromagnetism can be
observed. This is made possible by the collectively enhanced
magnetic moments of the condensate which in turn enhances magnetic
dipole-dipole interaction between different lattice sites.

In future work it will be interesting to study the dynamical
response of the system under the effect of an external
time-dependent longitudinal magnetic field. The instantaneous
magnetization in this case is likely to form hysteresis loops
which might find applications e.g. in quantum information
processing. Further studies should also include the properties of
finite temperature excitations of the system, which correspond to
spin waves. The extension of this work beyond one-dimensional
lattices also shows much promise: Although it is predominantly
ferromagnetic in 1D, the dipole-dipole interaction is anisotropic
and contains both ferromagnetic and antiferromagnetic terms. In
higher-dimensional lattices, the ground-state spin structure will
thus become much richer.

We conclude by remarking that in addition to this intrinsic
interest, the study of the magnetic properties of spinor
condensate lattices provide us with a highly controllable test
system to study fundamental static and dynamical aspects of
magnetism and lattice dynamics, including, e.g., the role of
dimensionality in phase transitions and macroscopic quantum
tunneling of the magnetic moments.

This work is supported in part by the US Office of Naval Research
under Contract No. 14-91-J1205, by the National Science Foundation
under Grant No. PHY98-01099, by the US Army Research Office, by
NASA, and by the Joint Services Optics Program.


\begin{references}

\bibitem{huang}See for example, Kerson Huang, {\em Statistical Mechanics}
(John Wiley \& Sons, New York, 1987).

\bibitem{heinzen} D. Heinzen, private communication (2001).

\bibitem{green}J. P. Burke, Jr. and J. L. Bohn, Phys. Rev. A {\bf 59},
1303 (1999).

\bibitem{yale1}B. P. Anderson and M. A. Kasevich, Science {\bf 281}, 1686
(1998).

\bibitem{zoller}D. Jaksch {\em et al.}, Phys. Rev. Lett. {\bf 81},
3108 (1998).

\bibitem{raizen} M. Raizen, private communication (2001)

\bibitem{carretero} R. Carretero-Gonzalez and K. Promislow, e-print
cond-mat/0105600, (2001).

\bibitem{mit}J. Stenger {\em et al.}, Nature (London) {\bf
396}, 345 (1998).

\bibitem{zhang}W. Zhang and D. F. Walls, Phys. Rev. A {\bf 57},
1248 (1998).

\bibitem{ho2}T. -L. Ho, Phys. Rev. Lett. {\bf 81}, 742 (1998).

\bibitem{ohmi}T. Ohmi and K. Machida, J. Phys. Soc. Jpn. {\bf 67}, 1882
(1998).

\bibitem{law}C. K. Law, H. Pu and N. P. Bigelow, Phys. Rev. Lett. {\bf 81},
5257 (1998).

\bibitem{solid}N. W. Ashcroft and N. D. Mermin, {\em Solid State Physics}
(Harcourt Brace College Publishers, New York, 1976).


\bibitem{yale2}C. Orzel {\em et al.}, Science, {\bf 291}, 2386 (2001).

\bibitem{hansch}M. Greiner {\rm et al.}, e-preprint cond-mat/0105105.

\bibitem{meacher}D. R. Meacher {\em et al.}, Phys. Rev. Lett. {\bf 74}, 1958
(1995).


\end{references}
\end{document}